\documentclass{article}

\usepackage{microtype}
\usepackage{graphicx}
\usepackage{subfigure}
\usepackage{booktabs} 
\usepackage{multirow}
\usepackage{booktabs}
\usepackage{array}   
\usepackage{makecell}
\usepackage{booktabs}
\usepackage{hyperref}

\usepackage[accepted]{icml2025}

\usepackage{amsmath}
\usepackage{amssymb}
\usepackage{mathtools}
\usepackage{amsthm}

\usepackage[capitalize,noabbrev]{cleveref}

\theoremstyle{plain}

\theoremstyle{definition}

\theoremstyle{remark}

\usepackage[textsize=tiny]{todonotes}

\icmltitlerunning{Self-Supervised Score-Based Despeckling for SAR Imagery via Log-Domain Transformation}

\begin{document}

\twocolumn[
\icmltitle{Self-Supervised Score-Based Despeckling for SAR Imagery \\ via Log-Domain Transformation}




\begin{icmlauthorlist}
\icmlauthor{Junhyuk Heo}{comp}
\end{icmlauthorlist}

\icmlaffiliation{comp}{TelePIX}

\icmlcorrespondingauthor{}{}

\icmlkeywords{Machine Learning, ICML}

\vskip 0.3in
]



\printAffiliationsAndNotice{} 

\begin{abstract}
The speckle noise inherent in Synthetic Aperture Radar (SAR) imagery significantly degrades image quality and complicates subsequent analysis. Given that SAR speckle is multiplicative and Gamma-distributed, effectively despeckling SAR imagery remains challenging. This paper introduces a novel self-supervised framework for SAR image despeckling based on score-based generative models operating in the transformed log domain. We first transform the data into the log-domain and then convert the speckle noise residuals into an approximately additive Gaussian distribution. This step enables the application of score-based models, which are trained in the transformed domain using a self-supervised objective. This objective allows our model to learn the clean underlying signal by training on further corrupted versions of the input data itself. Consequently, our method exhibits significantly shorter inference times compared to many existing self-supervised techniques, offering a robust and practical solution for SAR image restoration.
\end{abstract}

\section{Introduction}
\label{sec:introduction}

Synthetic Aperture Radar (SAR) is a highly important remote sensing technology, appreciated for its all-weather, day-and-night imaging capabilities essential for various Earth observation applications such as environmental monitoring~\cite{amitrano2021earth} and disaster management~\cite{moreira2013tutorial}. However, SAR images are intrinsically affected by speckle noise from coherent radar backscatter interference within each resolution cell~\cite{goodman1976some}. Speckle is multiplicative and is modeled by a Gamma distribution for intensity images\cite{perera2023sar}, with variance inversely proportional to the number of looks (\textit{L}). This noise significantly degrades image quality, disrupting automated analysis tasks such as segmentation and change detection. Consequently, effective despeckling is a critical preprocessing step in SAR image analysis.

To mitigate the negative effects of speckle, numerous despeckling techniques have been developed. Traditional algorithmic approaches, which operate without requiring clean ground truth (GT) images~\cite{frost1982model, argenti2002speckle, parrilli2011nonlocal, buades2005non, cozzolino2013fast, argenti2013tutorial}, are often computationally efficient but typically struggle to preserve fine image details while reducing noise. Addressing these limitations, deep learning methods emerged as a powerful alternative. In particular, supervised deep learning models tailored for SAR despeckling~\cite{chierchia2017sar, wang2017sar, zhang2017beyond} can outperform traditional methods when trained on large paired noisy-clean datasets. However, a major bottleneck for supervised learning in the SAR domain remains the scarcity of true speckle-free ground truth, as common approximations may not fully capture real SAR speckle characteristics, potentially limiting generalization on real-world data~\cite{dalsasso2020sar, vitale2021analysis}.

To circumvent the dependency on clean training data, self-supervised learning (SSL) strategies have evolved as a promising alternative for SAR despeckling. Prominent examples include methods like Noise2Noise~\cite{lehtinen2018noise2noise}, Noise2Void~\cite{krull2019noise2void}, Speckle2Void~\cite{molini2021speckle2void}, SAR-specific adaptations like SAR2SAR~\cite{dalsasso2021sar2sar} and MERLIN~\cite{dalsasso2024self}, and Deep Image Prior-based approaches such as S3DIP~\cite{albisani2025self}. While these SSL strategies innovatively reduce reliance on clean GT, some, like DIP-based methods including S3DIP, can be computationally intensive during inference due to their per-image optimization nature.

In this paper, we introduce a novel self-supervised framework for SAR image despeckling, leveraging score-based generative models ($\text{S}^4\text{DM}$) tailored to SAR speckle's unique characteristics. Our method directly addresses the multiplicative Gamma-distributed nature of speckle. It first transforms SAR image data via a logarithmic function to make the noise additive, followed by the Log-Yeo-Johnson(LYJ) transformation~\cite{yeo2000new, hu2024sar} to transform the residual noise into an approximately Gaussian distribution. We then train a neural network using a self-supervised objective inspired by the Corruption2Self methodology~\cite{tu2025score}. The network learns to predict a despeckled image from differently corrupted versions of the same noisy input, effectively learning the score function in the transformed domain. This approach requires no external clean reference images, offers robustness through direct learning from noisy SAR data. Our transformation-based, self-supervised score-matching approach $\text{S}^4\text{DM}$, demonstrates superior despeckling performance compared to baseline methods, while achieving inference times comparable to traditional algorithm-based techniques.


\section{Related Works}
\label{sec:related_works}


\paragraph{Score-Based Generative Models}
Score-based generative models represent a powerful and distinct paradigm in generative modeling and image restoration. These methods originate from the concept of score matching~\cite{hyvarinen2005estimation}, which aims to learn the score function—the gradient of the data's log-probability density. A key development, Denoising Score Matching (DSM)~\cite{vincent2011connection}, established that training a model to denoise data corrupted by Gaussian noise is equivalent to learning the score of the perturbed data distribution. This principle is fundamental to modern Denoising Diffusion Probabilistic Models (DDPMs)~\cite{ho2020denoising, song2020score}, which achieve high-fidelity image generation by iteratively reversing a noise process using the learned score. Crucially for our self-supervised approach, recent advancements have enabled the learning of score functions directly from noisy data, without clean targets. Techniques like Noise2Score~\cite{kim2021noise2score} and frameworks such as Corruption2Self~\cite{tu2025score} leverage principles like Tweedie's formula to achieve this. Our proposed method builds upon these self-supervised score estimation techniques, adapting them to the multiplicative speckle noise in SAR images through our specific noise transformation pipeline.

\begin{figure}[t]{\includegraphics[width=1.0\linewidth]{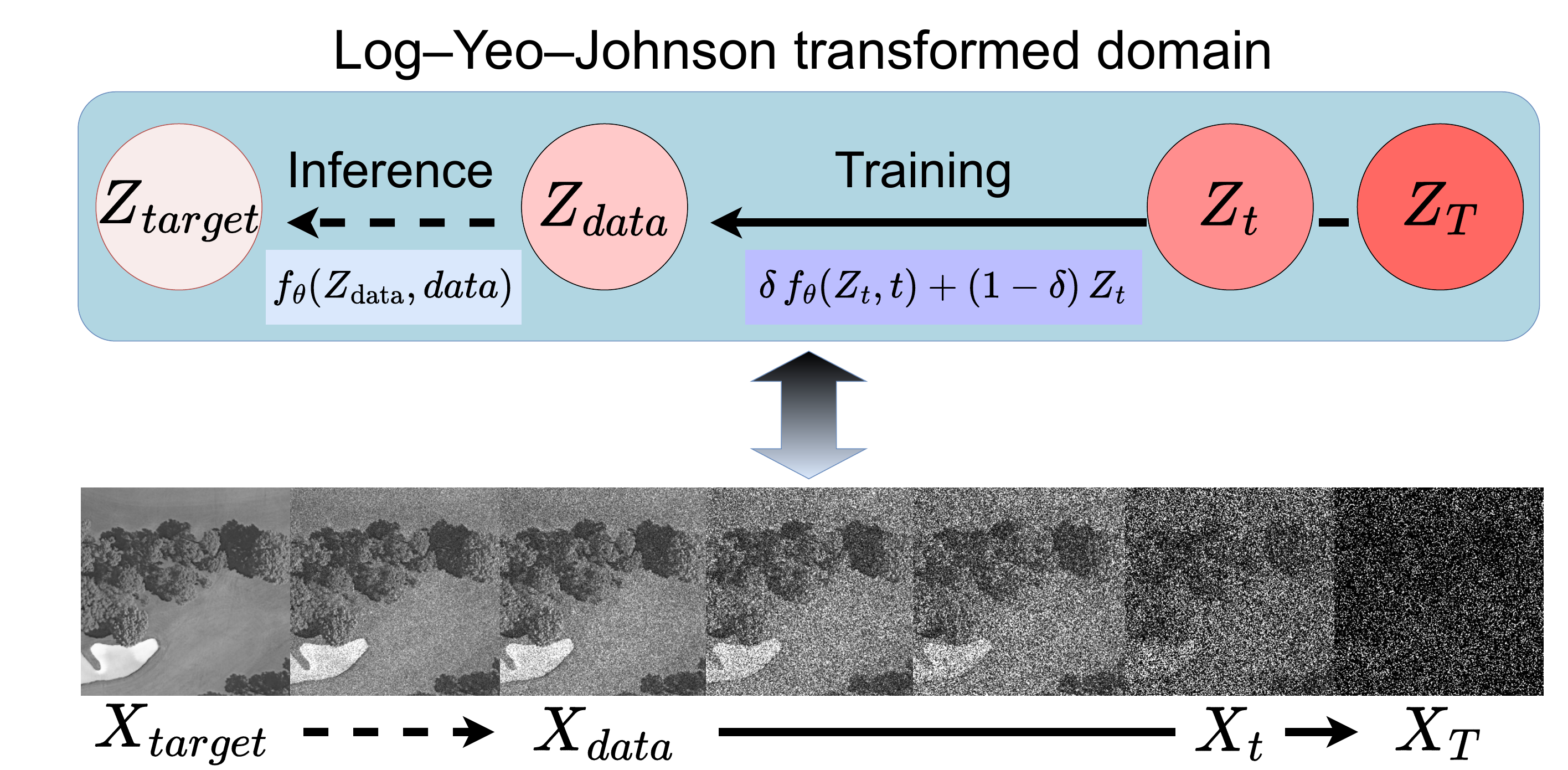}}\\
\caption{Overall framework of the proposed self-supervised SAR despeckling method ($\text{S}^4\text{DM}$). The diagram illustrates the transformation to the Log-Yeo-Johnson domain, the self-supervised training loop involving data corruption and reconstruction, and the inference path to obtain the despeckled image $X_{target}$.}
\label{fig:thumbnail} 
\end{figure}

\section{Method}
\label{sec:method}
This section details our score-based generative model for Synthetic Aperture Radar (SAR) despeckling. Our overall framework, illustrated in Figure~\ref{fig:thumbnail}, encompasses both the training and inference stages. The core strategy to handle multiplicative Gamma speckle involves employing LYJ transformation~\cite{yeo2000new} for noise normalization, which then allows for a self-supervised training objective inspired by Corruption2Self~\cite{tu2025score} in the transformed domain.

\subsection{Denoising Principle and Score Estimation}
\label{sec:denoising_principle}
The main goal of neural network $f_\theta(X_t, t)$ parameterized by $\theta$, is to estimate the clean image $X_0$ from a corrupted observation $X_t$ at noise level $t$, by computing the conditional expectation:
\begin{equation}
f_\theta(X_t, t) = \mathbb{E}[X_0|X_t]. \label{eq:cond_exp_general}
\end{equation}
For Gaussian noise, this expectation is directly related to the score function via Tweedie's formula:
\begin{equation}
\mathbb{E}[X_0|X_t] = X_t + \sigma_t^2 \nabla_{X_t} \log p_t(X_t), \label{eq:tweedie_general}
\end{equation}
where $\nabla_{X_t} \log p_t(X_t)$ is the score function of the data distribution $p_t(X_t)$, and $\sigma_t$ represents the standard deviation of the Gaussian noise added at time step $t$.

\subsection{Speckle Normalization via LYJ Transform}
\label{sec:lyj}

SAR images are affected by multiplicative Gamma-distributed speckle noise. This noise model can be expressed as:
\begin{equation}
X_{data} = X_{target} \odot \nu,
\label{eq:speckle_model}
\end{equation}
where $X_{data}$ is the observed noisy SAR amplitude, $X_{target}$ represents the underlying clean signal, $\nu$ is the speckle noise following a Gamma distribution $\nu \sim \Gamma(L,1/L)$ with $L$ being the number of looks, and $\odot$ denotes element-wise multiplication. This multiplicative is problematic for conventional denoising algorithms that assume additive Gaussian noise.

A logarithmic transform is first applied to convert the multiplicative noise into an additive model:
\begin{equation}
Y_{data} = \log X_{data} = \log X_{target} + \log \nu.
\label{eq:log_transform}
\end{equation}
While the noise becomes additive, the resulting log-Gamma noise ($\log \nu$) is still non-Gaussian. To address this, we employ the LYJ transformation $\mathcal{T}_{\lambda}$ to transform into an approximately Gaussian distribution, making them amenable to score-based modeling:
\begin{equation}
\mathcal{T}_{\lambda}(u) =
  \begin{cases}
    \dfrac{(u+1)^{\lambda}-1}{\lambda}, & \lambda\neq 0,\;u\ge 0,\\[4pt]
    \log(u+1), & \lambda=0,\;u\ge 0,\\[4pt]
    -\dfrac{(-u+1)^{2-\lambda}-1}{2-\lambda}, & \lambda\neq 2,\;u<0,\\[4pt]
    -\log(-u+1), & \lambda=2,\;u<0,
  \end{cases}
\label{eq:lyj}
\end{equation}
The parameter $\lambda^\dagger$, is selected by minimizing the sample kurtosis and skewness of the transformed noise~\cite{hu2024sar}. This process yields a transformed signal $Z_{data}$ where the noise component $\xi$ is approximately Gaussian:
\begin{equation}
  Z_{data} := \mathcal{T}_{\lambda^\dagger}(Y_{data})
     = \mathcal{T}_{\lambda^\dagger}\!\bigl(\log X_{target}\bigr)
       + \xi.
\label{eq:gauss_residual}
\end{equation}
Here, the residual noise $\xi$ is sampled as $\xi \sim \mathcal{N}(0,\sigma_{data}^2 I)$. This step is crucial as it allows the principled application of score-matching techniques and Tweedie's formula in the transformed domain.

\subsection{Self-Supervision via Data Corruption}
\label{sec:z_corruption}
After the SAR data transformed into $Z_{data}$ where noise is approximately Gaussian, we train the network $f_\theta$ using a self-supervised approach inspired by Corruption2Self. Let $Z_{target}$ denote the transformed version of the clean signal $X_{target}$. The noise components in $Z_{data}$ and $Z_{target}$ are assumed to be approximately Gaussian with variances $\sigma_{data}^2$ and $\sigma_{target}^2$ (where $\sigma_{target}^2 \approx 0$), respectively.

\paragraph{Gaussian Diffusion Step.}
To facilitate self-supervised learning, we further corrupt $Z_{data}$ by adding Gaussian noise, creating progressively noisier versions $Z_t$. For any noise variance $\sigma_t^2 > \sigma_{data}^2$:
\begin{equation}
  Z_t = Z_{data} + \sqrt{\sigma_t^{2}-\sigma_{data}^{2}}\;\xi, \qquad \xi \sim \mathcal N(0,I).
  \label{eq:z_t_from_data}
\end{equation}
Equivalently, $Z_t$ can also be expressed as a more corrupted version of $Z_{target}$:
\begin{equation}
  Z_t = Z_{target} + \sqrt{\sigma_t^{2}-\sigma_{target}^{2}}\;\xi', \qquad \xi' \sim \mathcal N(0,I).
  \label{eq:z_t_from_target}
\end{equation}

\paragraph{Two Tweedie Expectations.}
Applying Tweedie’s formula in Eq.~\eqref{eq:tweedie_general} to $Z_t$ in the forward processes described by Eqs.~\eqref{eq:z_t_from_data} and~\eqref{eq:z_t_from_target} gives the conditional expectations:
\begin{align}
  \mathbb E[Z_{{data}} \mid Z_t]
    &= Z_t + (\sigma_t^{2}-\sigma_{{data}}^{2}) \mathcal{S}_t, \label{eq:tweedie_data} \\
  \mathbb E[Z_{{target}} \mid Z_t]
    &= Z_t + (\sigma_t^{2}-\sigma_{{target}}^{2}) \mathcal{S}_t. \label{eq:tweedie_target}
\end{align}
Let $\mathcal{S}_t = \nabla_{Z_t}\!\log p_t(Z_t)$ denote the score term. Both expectations rely on the same score function $\mathcal{S}_t$ of the noisy data $p_t(Z_t)$.

\paragraph{Formulating the Self-Supervised Objective.}
Leveraging the common score term in Eqs.~\eqref{eq:tweedie_data} and~\eqref{eq:tweedie_target}, we can relate the conditional expectation of the observable $Z_{data}$ to that of the desired $Z_{target}$ without explicit knowledge of $X_{target}$ or the score function itself:
\begin{equation}
\label{eq:blend}
\begin{split}
  \mathbb{E}[Z_{data} \mid Z_t]
    &= \delta\,\mathbb{E}[Z_{{target}} \mid Z_t] + (1-\delta)\,Z_t,
\end{split}
\end{equation}
where $\delta := ({\sigma_t^{2}-\sigma_{{data}}^{2}}) / ({\sigma_t^{2}-\sigma_{{target}}^{2}})$. This blending equation is key to our self-supervised objective, following the Corruption2Self methodology.

\begin{table*}[t]
\centering
\footnotesize 
\setlength{\tabcolsep}{4pt} 
\caption{Comparison of ENL / inference time in seconds for different methods on Sentinel-1 GRD test images. 'Img. No.' denotes Image Number. The S3DIP+lf method indicates the S3DIP model trained using a loss fusion technique. Higher ENL values denote stronger speckle reduction, while lower times indicate faster processing. \textbf{Bold} values indicate the best performance, and \underline{underlined} values indicate the second best. ENL Values are averaged over four homogeneous ROIs per image.}
\label{tab:quantitative_results}
\vskip 0.15in 

\begin{tabular}{@{} >{\centering\arraybackslash}m{2cm} >{\centering\arraybackslash}m{1.5cm} cccccc@{}} 
\toprule
Category & Img. No. & \multicolumn{2}{c}{\textbf{Algorithm-based}} & \multicolumn{4}{c}{\textbf{Deep Learning-based}} \\
\cmidrule(lr){3-4} \cmidrule(lr){5-8}
& & FANS & SAR-BM3D & DIP & S3DIP & S3DIP+lf & \textbf{$\text{S}^4\text{DM}$}\textbf{(Ours)} \\
\midrule
\multirow{2}{*}[-0.5ex]{Agricultural}
& Image 1 & 31.70 / \textbf{0.01} & \underline{35.60} / 0.52 & 16.18 / 26.76 & 30.56 / 84.12 & 34.58 / 84.77 & \textbf{49.31} / \underline{0.05} \\
& Image 2 & 68.32 / \textbf{0.01} & 121.02 / 0.52 & 41.26 / 35.83 & \underline{144.09} / 81.07 & 88.16 / 81.50 & \textbf{217.22} / \underline{0.05} \\
\midrule
\multirow{2}{*}[-0.5ex]{Mountain}
& Image 3 & \textbf{55.71} / \textbf{0.01} & 27.92 / 0.52 & 12.71 / 37.14 & 22.03 / 77.72 & 36.29 / 79.45 & \underline{43.85} / \underline{0.05} \\
& Image 4 & \underline{26.51} / \textbf{0.01} & 23.38 / 0.52 & 16.73 / 35.82 & 24.41 / 79.03 & 25.58 / 81.30 & \textbf{27.85} / \underline{0.05} \\
\bottomrule
\end{tabular}
\vskip -0.1in 
\end{table*}

\begin{figure*}[t]

\subfigure[Sentinel-2]{\includegraphics[width=0.13\linewidth]{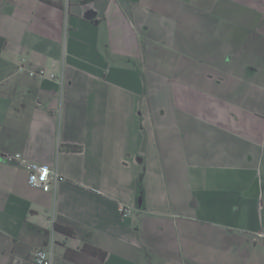}}\hfill
\subfigure[Sentinel-1]{\includegraphics[width=0.13\linewidth]{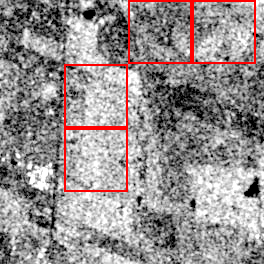}}\hfill
\subfigure[FANS]{\includegraphics[width=0.13\linewidth]{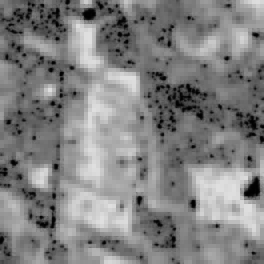}}\hfill
\subfigure[SAR-BM3D]{\includegraphics[width=0.13\linewidth]{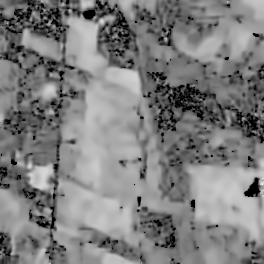}}\hfill
\subfigure[DIP]{\includegraphics[width=0.13\linewidth]{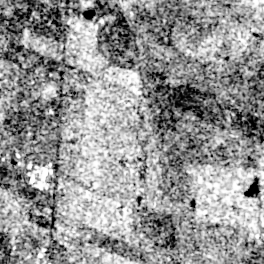}}\hfill
\subfigure[S3DIP]{\includegraphics[width=0.13\linewidth]{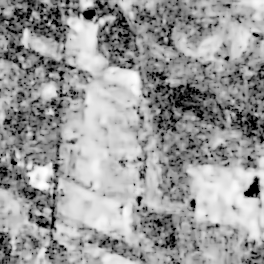}}\hfill
\subfigure[\textbf{$\text{S}^4\text{DM}$(Ours)}]{\includegraphics[width=0.13\linewidth]{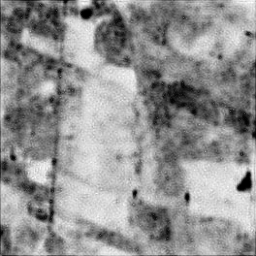}}\\
\caption{Despeckling results for Agricultural area Image 1, from Table~\ref{tab:quantitative_results}. (a) Sentinel-2 optical reference. (b) Sentinel-1 input. (c)-(g) Outputs from FANS, SAR-BM3D, DIP, S3DIP, and $\text{S}^4\text{DM}$ (Ours). See Appendix~\ref{app:additional_qualitative_results} for results on all benchmark dataests.}
\label{fig:qualitative_results_image} 
\end{figure*}

\subsection{Training Objective}
\label{sec:training_objective}
The neural network $f_\theta(Z_t,t)$ is trained to output $\widehat{Z}$, an estimate of $\mathbb E[Z_{{target}} \mid Z_t]$. As $\sigma_{{target}}^2 \approx 0$, this means $f_\theta$ aims to recover the transformed clean signal $\mathcal{T}_{\lambda^\dagger}(\log X_{target})$.
The training objective is to minimize the difference between $Z_{data}$ and its reconstruction based on the network's output $\widehat{Z}$ and $Z_t$, using Eq.~\eqref{eq:blend}. The weighted self-supervised loss is:
\begin{equation}
  \mathcal L(\theta)
    = \mathbb E_{Z_{data},t}
      \bigl[
        w(t) \Vert
          \delta\,\widehat{Z} + (1-\delta)\,Z_t - Z_{data}
        \Vert_2^{2}
      \bigr],
  \label{eq:selfsup_loss}
\end{equation}
where $w(t)$ serves as a weighting term that modulates the influence of samples from different noise levels $t$. This term is defined as a function of the noise variance $\sigma^2_t$. By minimizing this loss, $f_\theta(Z_t, t)$ (which produces $\widehat{Z}$) is trained such that the term $\delta\,\widehat{Z} + (1-\delta)\,Z_t$ accurately approximates $\mathbb{E}[Z_{data}|Z_t]$. Given the fixed relationship in Eq.~\eqref{eq:blend}, this implicitly drives $\widehat{Z}$ to be a robust estimate of $\mathbb{E}[Z_{target}|Z_t]$.

\subsection{Inference}
\label{sec:inference}
For inference, the observed noisy SAR image $X_{{data}}$ is transformed to $Z_{data} = \mathcal{T}_{\lambda^\dagger}(\log X_{data})$. The trained model $f_\theta$ then predicts the denoised transformed signal $\widehat Z = f_\theta(Z_{data}, data)$, using the estimated noise level $\sigma_{{data}}$ of the input $Z_{data}$. The final despeckled SAR amplitude $\widehat X$ is recovered by inverting the LYJ and logarithmic transformations:
\begin{equation}
  \widehat X
    =
    \exp(
        \mathcal{T}_{\lambda^\dagger}^{-1}(\widehat Z)
    ).
    \label{eq:inference}
\end{equation}

\section{Experiments}
\label{sec:experiments}
This section validates our proposed SAR despeckling method. We describe the experimental setup, evaluation metrics, and then present and discuss the comparative results.

\subsection{Experimental Setup}

\paragraph{Dataset and Preprocessing.}
We use 10,000 Sentinel-1 Ground Range Detected (GRD) amplitude images (derived from VV/VH polarizations of the SSL4EO-S12 dataset \cite{Wang2023ssl4eo}) for training our model.

\paragraph{Baseline Methods}
To benchmark the performance of our proposed method ($\text{S}^4\text{DM}$), we selected baseline models specifically capable of processing and learning from Sentinel-1 GRD amplitude data. These established baselines are broadly categorized into algorithm-based methods, including traditional techniques such as FANS~\cite{cozzolino2013fast} and SAR-BM3D~\cite{parrilli2011nonlocal}, and representative deep learning-based approaches adapted for SAR despeckling, such as Deep Image Prior (DIP)~\cite{ulyanov2018deep} and S3DIP~\cite{albisani2025self}. Notably, the S3DIP method can be enhanced with a loss fusion technique, the details of which are described in Appendix~\ref{app:additional_experiments_details}. These selected methods serve as points of comparison for evaluating the despeckling performance and computational efficiency presented in Table~\ref{tab:quantitative_results} and Figure~\ref{fig:qualitative_results_image} and Appendix~\ref{app:additional_qualitative_results}.

\subsection{Evaluation Approach}
Performance is assessed qualitatively through visual inspection of the despeckled images. Representative examples are presented in Figure~\ref{fig:qualitative_results_image}, with further results available in Appendix~\ref{app:additional_qualitative_results}. For quantitative assessment, we use the Equivalent Number of Looks (ENL). ENL serves as a standard no-reference metric to quantify speckle reduction, particularly in homogeneous image regions, where higher values indicate more effective speckle suppression and improved smoothness~\cite{anfinsen2008estimation}. A detailed description of our ENL calculation methodology, which includes the criteria for selecting Regions of Interest (ROIs), is provided in Appendix~\ref{app:additional_experiments_details}.

\subsection{Results and Discussion}

\paragraph{Visual Analysis}
Figure~\ref{fig:qualitative_results_image} presents qualitative despeckling results for "Image 1." Both traditional algorithm-based methods and other deep learning-based baselines demonstrate varying degrees of speckle reduction, sometimes with trade-offs in detail preservation or the introduction of artifacts. In contrast to these baselines, our proposed method, $\text{S}^4\text{DM}$ (Figure~\ref{fig:qualitative_results_image}(g)), achieves strong speckle suppression while preserving key details and textures, yielding cleaner images (see Appendix~\ref{app:additional_qualitative_results} for more examples).

\paragraph{Quantitative ENL Results}
Table~\ref{tab:quantitative_results} summarizes the quantitative performance, measured by the average ENL over selected homogeneous ROIs from four test images. As indicated in Table~\ref{tab:quantitative_results}, our proposed method, $\text{S}^4\text{DM}$, achieves higher ENL scores compared to the baseline methods across all tested images. This suggests a superior capability of $\text{S}^4\text{DM}$ in smoothing speckle noise effectively in homogeneous regions.

\section{Conclusion}
\label{sec:conclusion}
We introduced $\text{S}^4\text{DM}$, a novel self-supervised, score-based framework for despeckling SAR imagery. By employing a LYJ transformation pipeline to normalize speckle noise and a Corruption2Self-inspired training objective, $\text{S}^4\text{DM}$ effectively reduces speckle and preserves image details without requiring clean reference data, offering a practical solution for SAR image restoration.
\nocite{langley00}

\bibliography{ref}
\bibliographystyle{icml2025}

\newpage
\appendix
\onecolumn
\section{Additional Experiments resutls}
\label{app:additional_qualitative_results}
\begin{figure*}[ht!]
\subfigure[Sentinel-2]{\includegraphics[width=0.13\linewidth]{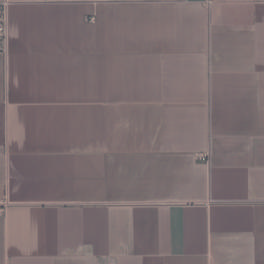}}\hfill
\subfigure[Sentinel-1]{\includegraphics[width=0.13\linewidth]{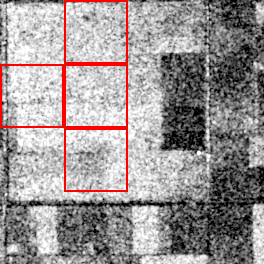}}\hfill
\subfigure[FANS]{\includegraphics[width=0.13\linewidth]{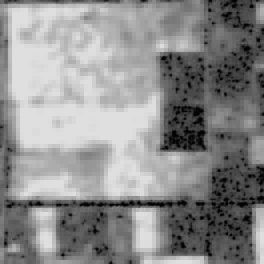}}\hfill
\subfigure[SAR-BM3D]{\includegraphics[width=0.13\linewidth]{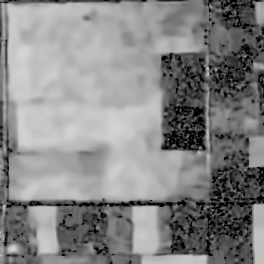}}\hfill
\subfigure[DIP]{\includegraphics[width=0.13\linewidth]{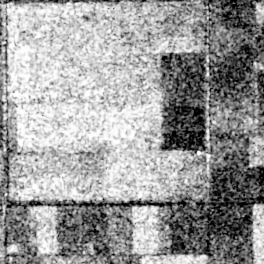}}\hfill
\subfigure[S3DIP]{\includegraphics[width=0.13\linewidth]{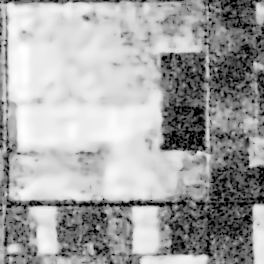}}\hfill
\subfigure[\textbf{$\text{S}^4\text{DM}$(Ours)}]{\includegraphics[width=0.13\linewidth]{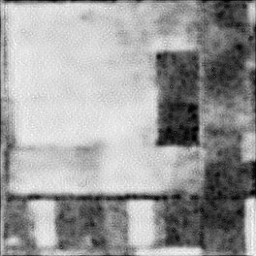}}\\

\subfigure[Sentinel-2]
{\includegraphics[width=0.13\linewidth]{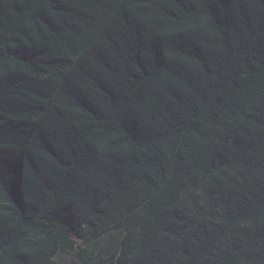}}\hfill
\subfigure[Sentinel-1]
{\includegraphics[width=0.13\linewidth]{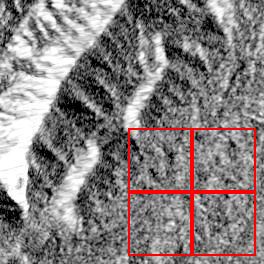}}\hfill
\subfigure[FANS]{\includegraphics[width=0.13\linewidth]{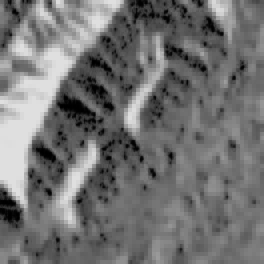}}\hfill
\subfigure[SAR-BM3D]{\includegraphics[width=0.13\linewidth]{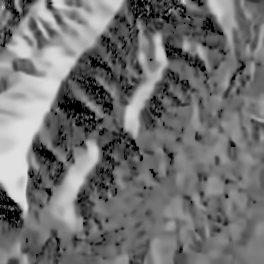}}\hfill
\subfigure[DIP]{\includegraphics[width=0.13\linewidth]{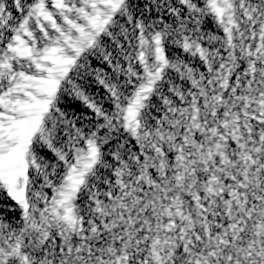}}\hfill
\subfigure[S3DIP]{\includegraphics[width=0.13\linewidth]{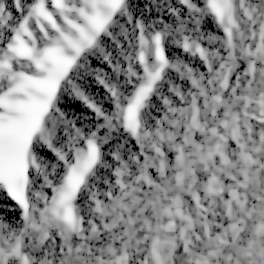}}\hfill
\subfigure[\textbf{$\text{S}^4\text{DM}$(Ours)}]{\includegraphics[width=0.13\linewidth]{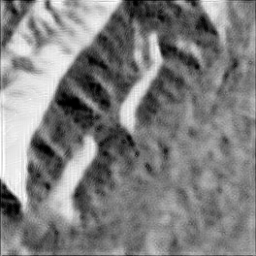}}\\

\subfigure[Sentinel-2]
{\includegraphics[width=0.13\linewidth]{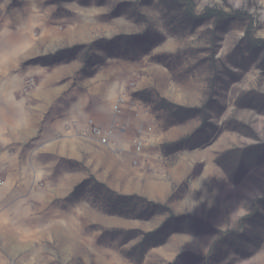}}\hfill
\subfigure[Sentinel-1]
{\includegraphics[width=0.13\linewidth]{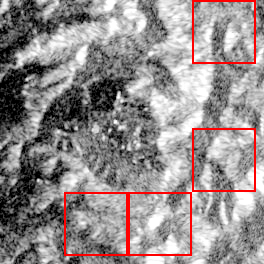}}\hfill
\subfigure[FANS]{\includegraphics[width=0.13\linewidth]{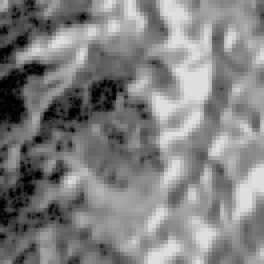}}\hfill
\subfigure[SAR-BM3D]{\includegraphics[width=0.13\linewidth]{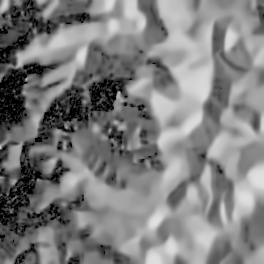}}\hfill
\subfigure[DIP]{\includegraphics[width=0.13\linewidth]{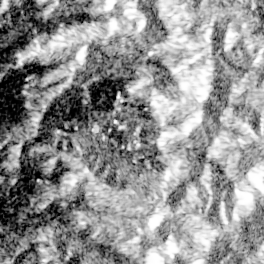}}\hfill
\subfigure[S3DIP]{\includegraphics[width=0.13\linewidth]{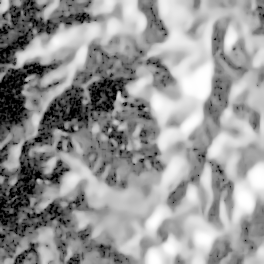}}\hfill
\subfigure[\textbf{$\text{S}^4\text{DM}$(Ours)}]{\includegraphics[width=0.13\linewidth]{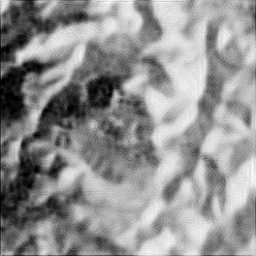}}\\

\caption{Additional qualitative despeckling results for test images (Image 2, Image 3, and Image 4), supplementing the quantitative metrics presented in Table~\ref{tab:quantitative_results}. 
Subfigures (a)-(g) display the results for Agricultural area Image 2. 
Subfigures (h)-(n) display the results for Mountain area Image 3. 
Subfigures (o)-(u) display the results for Mountain area Image 4. 
For each image, the sequence shown is: (a) Sentinel-1 Input, followed by the outputs from (b) FANS~\cite{cozzolino2013fast}, (c) SAR-BM3D~\cite{parrilli2011nonlocal}, (d) DIP~\cite{ulyanov2018deep}, (e) S3DIP~\cite{albisani2025self}, (f) S3DIP+lf~\cite{albisani2025self}, and (g) \textbf{$\text{S}^4\text{DM}$ (Ours)}.}
\label{fig:appendix_qualitative_results}
\end{figure*}

\section{Additional Experiments Details}
\label{app:additional_experiments_details}

\subsection{Implementation Details}
All models, including our proposed method ($\text{S}^4\text{DM}$) and the deep learning baselines, are trained and evaluated on a single NVIDIA RTX 4090 GPU. For a fair comparison of performance and convergence characteristics under similar computational constraints, training for our method and all compared self-supervised baselines is conducted for a fixed 10,000 iterations.

\subsection{ENL Calculation Methodology}
For the quantitative evaluation using the ENL, test images are selected from two primary terrain categories: "agricultural" and "mountain", with two distinct images chosen from each category to assess performance across different scene types. The ENL for each despeckled image is calculated as follows:
\begin{enumerate}
    \item The image is divided into non-overlapping patches of a predefined size (e.g., 32x32 pixels).
    \item For each patch, the variance of pixel intensities is computed.
    \item Four patches exhibiting the lowest variance (i.e., considered the most homogeneous) are automatically selected as the Regions of Interest (ROIs). An example of such selected ROIs is visually indicated by red boxes overlaid on the input Sentinel-1 image in Figure~\ref{fig:qualitative_results_image}(b) in the main paper.
    \item The ENL is then computed for each of these four ROIs using the formula ENL = $(\mu_I / \sigma_I)^2$, where $\mu_I$ is the mean and $\sigma_I$ is the standard deviation of pixel intensities within the ROI.
    \item The final reported ENL score for an image is the average of the ENL values obtained from these four selected ROIs.
\end{enumerate}
This procedure ensures a consistent and objective approach to selecting homogeneous regions for ENL calculation, providing a robust measure of speckle reduction.

\subsection{S3DIP Loss Fusion Technique}
The S3DIP method \cite{albisani2025self}, an extension of Deep Image Prior (DIP) \cite{ulyanov2018deep} for SAR despeckling, can optionally incorporate a "loss fusion" mechanism. This involves adding a guidance term to its primary loss function. This guidance term encourages the output of S3DIP to align with results from established model-based denoisers like FANS \cite{cozzolino2013fast} and SAR-BM3D \cite{parrilli2011nonlocal}. Essentially, the total loss becomes a weighted combination of a reconstruction loss, a statistical loss for the speckle, and this guidance loss. The variant using this approach is denoted as S3DIP+lf in our experiments.



\end{document}